# Modelling and Understanding of Highly Energy Efficient Fluids

R.J.K.A. Thamali, A. A. KKumbalatara, D. D. Liyanage, Ajith Ukwatta, JinasenaHewage, SanjeevaWitharana

**Abstract:** Conventional heat carrier liquids have demonstrated remarkable enhancement in heat and mass transfer when nanoparticles were suspended in them. These liquid-nanoparticle suspensions are now known as *Nanofluids*. However the relationship between nanoparticles and the degree of enhancement is still unclear, thus hindering the large scale manufacturing of them. Understanding of the energy and flow behaviour of nanofluids is therefore of wide interest in both academic and industrial context. In this paper we first model the heat transfer of a nanofluid in convection in a circular tube at macro-scale by using CFD code of OpenFoam. Then we zoon into nano-scale behaviour using the Molecular Dynamics (MD) simulation. In the latter we considered a system of water and Gold nanoparticles. A systematic increase of convective heat transfer was observed with increasing nanoparticle concentration. A maximum enhancement of 7.0% was achieved in comparison to base fluid water. This occurred when the gold volume fraction was 0.015. Subsequent MD simulations confirmed a dense water layer formed 2nm away from the gold nanoparticle boundary. It is speculated that the observed increase in thermal conductivity was principally caused by this dense water layer formed around each nanoparticle.

## 1. Nomenclature

| Symbol | Description (unit) |
|---|---|
| $\rho$ | Density (kg/m$^3$) |
| U | Velocity vector (m/s) |
| P | Pressure (kg/m/s$^2$) |
| $\mu$ | Dynamic viscosity (kg/m/s) |
| C | Specific heat capacity (J/kg/K) |
| k | Thermal conductivity (W/m/K) |
| $\varphi$ | Volume fraction |
| h | Convective heat transfer coefficent (W/m$^2$/K) |
| Nu | Nusselt number |
| $V_{ij}^{pair}$ | Repulsive potential between atom cores (kJ/mol) |
| $\rho_i$ | Interaction between valence electrons and cores (kJ/mol) |
| $\alpha_{ij}$ | Minimum potential energy between two atoms (kJ/mol) |
| $\epsilon_{ij}, \epsilon_{ii}, c_i$ | Parameters use to tune the potential. |
| $D_{ij}$ | Well depth (angstrom) |
| $r^0$ | Minimum potential distance (angstrom) |
| $\beta_{ii}$ | Controls the 'width' of the potential |

## 2. Introduction

Heat Transfer is one of most important processes in all areas of engineering, from power plants to microelectronics. At industrial scale, heat transfer modes are dominated by forced convection, followed by boiling and condensation. Moving a fluid requires pumping power and costs electricity. Despite efficient heat transfer mechanisms introduced during the last half-century, they have fallen short of matching the ever increasing heat loads. A breakthrough is urgently needed.

*Nanofluids* are the blends of nanometer-sized particles and heat transfer liquids. Typically the particle sizes are below 100nm. They were introduced in mid-1990s as highly thermal conductive fluids [1]. Metals are higher in thermal conductivity than water. When they are blended together, the resultant thermal conductivity should naturally be a larger than that of water. People have earlier used micrometer and millimetre size particles to increase thermal conductivity of liquids. These large particles however introduced several difficulties such as,

- settling down faster due to their weight.
- causing erosion in pipe lines and clogging filters.
- high pressure drop due to increase of friction

Nanoparticles, due to their extremely small sizes, did not create these problems. Most important aspect however was, the thermal conductivity of nanofluids largely surpassed the calculated value [2]. Tests with other liquids, for example glycols and engine oils etc., behaved in similar style. Soon the nanofluids were examined for convective heat transfer. There too, the improvement was remarkable [3]. The smart fluid sought after for a long time had possibly arrived. But there was one problem; what is the relationship between the nanofluid's thermal properties and the nanoparticle concentration? Once this is solved, the smart fluids will be ready to be manufactured. First part of this paper presents macro-scale modelling of thermal conductivity of nanofluids using the CFD code of OpenFOAMversion 2.3.1.

There are four mechanisms widely speculated to be contributing to enhanced thermal behaviour of nanofluids. They are [4]; Brownian motion, Aggregation,Mode of heat flow and Liquid layering around,the nanoparticles.

Researchers further investigated the liquid layering around nanoparticles [5]. Initial simulation results showed this was a possible mechanism. However their work fell short of arriving at a convincing conclusion. More recently Shin and Banerjee [6] attributed the unexpectedly high enhancement in thermal conductivity to three reasons;

a. Surface energy of the boarder atoms of the Nano-particle is higher due to low vibration and higher amplitudes of the vibration.
b. Interaction between nanoparticle and liquid molecules make additional thermal storage area around the nanoparticle.
c. Liquid layering enhances the thermal conductivity due to enhancement of specific heat. This happen because of the shorter inter-molecular mean free path.

Going on this line, the second part of this paper presents the Molecular Dynamic simulation of liquid layering around a nanoparticle in a liquid. For this we used OpenMD version 2.2 on a gold-water system.

## 3. Literature Review

It is clear that better understanding of Nano-Fluid make significant changes in the engineering applications. Such as boilers, vehicle radiators, chill plants and many more. Therefore it is better to review the existing studies conducted by researches to understand about the properties of Nano-Fluids. Study about thermal conductivity of Nano-Fluids was started in 1993. Research works had been conducted in experimental, empirical and numerical to study about the properties in Nano-Fluids.

Table 1 shows some of the experimental data gathered by researches relevant to the thermal conductivity in Nano-fluids. It was evident that the thermal conductivity of Nano-fluid had significantly increased in some instances when it is compared to thermal conductivity in base fluid.

**Table 1: experimental investigation of enhancement in Nano-fluids**

| Nano-Fluid composition | Results |
|---|---|
|  |  |

*Eng. R.J.K.A.Thamali, M.Sc,B.Sc. Eng. (Moratuwa)*
*Eng. A.A.K.Kumbalathara, B.Sc. Eng. (Ruhuna)*
*Eng. D.D.Liyanage, B.Sc. Eng. (Ruhuna)*
*Eng.Ajith Ukwatta, B.Sc. Eng. (Moratuwa)*
*Dr.Jinasena W.Hewage, PhD(USA), B.Sc. Sci. (Ruhuna)*
*Dr.Sanjeeva Witharana, PhD (UK), M. Sc(Sweden), B.Sc.(Moratuwa)*

| | |
|---|---|
| Al₂O₃– Water [7] | 41% enhancement of heat transfer coefficient at 2.5% Volume of nanoparticle |
| Graphite- water [8] | 22% enhancement of heat transfer coefficient at 2.5Vol.% nanoparticles |
| titanate/water [9] | remarkable increasing of convective heat transfer with increasing of nanoparticles concentration and aspect ratio |
| Al₂O₃ – water [10] | Increasing of heat transfer in the electronic cooling system up to 40% at 6.8 Vol.% of nanoparticles |
| CNT – water [11] | Remarkable enhancement of heat transfer coefficient |
| Au – water [12] | 8.3% enhancement at 0.026Vol% |

## 4. Materials and Methods

**Mathematical modelling**

Nanofluid was treated as a homogeneous single phase fluid. Fluid flow is considered to be laminar and heat transferring method is forced convection. Flow and heat transfer of a nanofluid at steady state can be described as follows [13].

Mass conservation equation,
$$\nabla \cdot (\rho_{nf} U) = 0 \qquad (i)$$

Momentum conservation equation,
$$\nabla \cdot (\rho_{nf} UU) = -\nabla P + \nabla \cdot (\mu_{nf} \nabla U) \qquad (ii)$$

Energy conservation equation,
$$\nabla \cdot (\rho_{nf} UCT) = \nabla(k_{nf} \nabla T) \qquad (iii)$$

Density of the nanofluid is calculated using the equation below, with $\varphi$ as the volume fraction.
$$\rho_{nf} = (1 - \varphi)\rho_{bf} + \varphi \rho_{np} \qquad (iv)$$

Effective heat capacity of the nanofluid is given by,
$$C_{nf} = \frac{(1-\varphi)\rho_{bf} C_{bf} + \varphi \rho_{np} C_{np}}{\rho_{nf}} \qquad (v)$$

$$\mu_{nf} = \mu_{bf}(1 + 2.5\varphi) \qquad (vi)$$

Viscosity of the nanofluid is depending on several factors such as nanoparticle size and their concentration. Convective heat transfer coefficient has not shown an impact of the particle size for a given Reynolds Number. Therefore, this work did not consider about the particle size [14].

The liquid layer parameter will be introduced to the system of equations. Let $\beta$ be the ratio of the nano-layer thickness to the original particle radius. Due to this liquid layering, the thermal conductivity of the nanofluid can be written as follows,

$$k_{nf} = \left[\frac{k_p + 2k_{bf} + 2(k_p + k_{bf})(1+\beta^3)\varphi}{k_p + 2k_{bf} - 2(k_p - k_{bf})(1+\beta^3)\varphi}\right] k_{bf} \qquad (vii)$$

For the simulation of interaction between metal atoms in the nanoparticle, Sutton-Chen potential is used. [15]

$$U_{tot} = \sum_i \left[\frac{1}{2} \sum_{j \neq i} \epsilon_{ij} V_{ij}^{pair}(r_{ij}) - c_i \epsilon_{ii} \sqrt{\rho_i}\right] \qquad (viii)$$

Where $V_{ij}^{pair}$ is the repulsive potential between atom cores. $\rho_i$ is the interaction between valence electrons and cores of the atoms. $\epsilon_{ij}$, $\epsilon_{ii}$ and $c_i$ parameters use to tune the potential.

$$V_{ij}^{pair}(r) = \left(\frac{\alpha_{ij}}{r_{ij}}\right)^{r_{ij}} \qquad (ix)$$

$$\rho_i = \sum_{j=1} \left(\frac{\alpha_{ij}}{r_{ij}}\right)^{m_{ij}} \qquad (x)$$

In MDS there are several water models that can be used. Here the SPCE water model is used because its reliability is high. In SPCE water model it individually considers the attractions between Oxygen – Oxygen, Hydrogen – Hydrogen, and Oxygen – Hydrogen. The Potential between water molecules are set by Lennard Jones potential. [15]

$$V_{NB}(r) = 4\epsilon_{ij}\left(\left(\frac{\sigma_{ij}}{r}\right)^{12} - \left(\frac{\sigma_{ij}}{r}\right)^6\right) \qquad (xi)$$

The interaction between water and nanoparticle is modelled using two potentials. Now consider the SPCE water model in figure 1 below.

The interaction between nanoparticle molecules vs. SPCE water molecules can act in two ways; Oxygen atom vs. nanoparticle metal atom and Hydrogen atom vs nanoparticle metal atom.

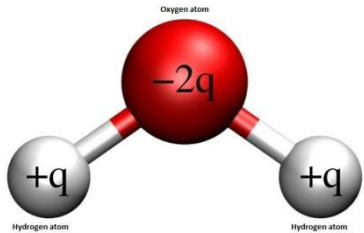

Figure 1: SPCE water model

The attraction between Oxygen atom and the nanoparticle atom are given by Shifted-Morse potential

$$V_{NB}(r) = D_{ij}(e^{-2\beta_{ii}(r-r^0)} - 2e^{-\beta_{ii}(r-r^0)}) \qquad \text{(xii)}$$

And the potential between Hydrogen atom and nanoparticle atom can be implanted using Repulsive-Morse potential.

$$V_{NB}(r) = D_{ij}(e^{-2\beta_{ii}(r-r^0)}) \qquad \text{(xiii)}$$

**CFD Simulation of Thermal Conductivity in Nano-Fluids**

Numerical simulation is carried out using open source CFD software OpenFOAM 2.3.1.[17,18] It was a water-Au nanofluid in single phase fully-developed laminar flow in a circular copper tube. The tube was 1mm thick and 6mm internal diameter. A constant wall heat flux of 1.8 W/cm² was applied to the tube exterior. It is further assumed the simulation fluid enters at 298K temperature and constant velocity. Figure 2 shows details of the arrangement.

Thermopysical properties of the nanofluid were calculated using equations (iv) to (vi). $\beta$ was assumed as 0.1 [5]. Au nanoparticle volume fractions were set to 0.002%, 0.01%, 0.015%, 0.02% and 0.025% and simulated at constant inlet velocity of 0.0706ms$^{-1}$. Convective heat transfer coefficient and Nusselt number were calculated using simulation results.

2D Mesh was generated using BlockMesh utility in OpenFOAM with mesh size of 1000×20. Conjugate heat transfer between the pipe material and fluid flow was simulated. Simulation results at steady state with different conditions were compared with experimental data in literature.

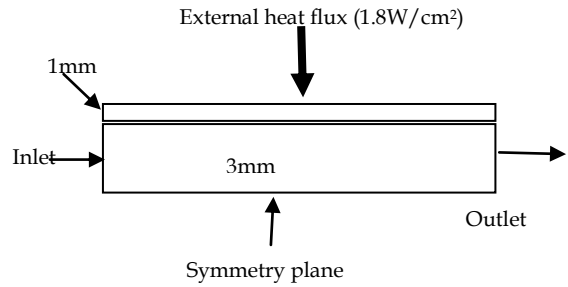

Figure 2: Schematic of the flow of nanofluid through the pipe

**MDS Simulation of liquid layering around a nanoparticle**

Molecular Dynamic Simulation was conducted using the open source software called OpenMD version 2.2. Simulation parameters are stated in Tables 2 through 5.

Table 2: Sutton-Chen potential parameters for Au [16]

| parameter | value |
|---|---|
| $\epsilon$ (eV) | 0.012896 |
| $\alpha$ (angstroms) | 4.08 |
| $m$ | 8.0 |
| $n$ | 10.0 |
| $c$ | 34.428 |

Table 3: Lennard Jones potential for Oxygen atom in SPCE water model [19]

| parameter | value |
|---|---|
| $\epsilon$(eV) | 0.15532 |
| $\sigma$ | 3.16549 |

Table 4: Shifted-Morse potential Oxygen atom in SPCE water model Vs Au nanoparticle [15]

| parameter | value |
|---|---|
| $r^0$ | 3.70 |
| $\beta$ | 0.769 |
| $D$ | 0.0424 |

**Table 5: Repulsive-Morse potential Hydrogen atom in SPCE water model Vs Au nanoparticle [15]**

| parameter | value |
|---|---|
| $r^0$ | -1.00 |
| $\beta$ | 0.769 |
| $D$ | 0.00850 |

The simulation domain consisted of 4nm gold nanoparticle and 1700 water molecules around it.

## 5. Results and Discussion

**CFD Simulation of nanofluid**

Generated mesh and simulated results for 0.01% and 0.02% Au concentration nanofluids are shown in figures 3, 4 and 5 respectively. Convective heat transfer coefficients and Nusselt numbers were calculated from the simulation results for nanofluids with different particle volume fractions.

Variation of the convective heat transfer coefficient ($h$) with respect to nanoparticle concentration ($\varphi$) of the nanofluid is shown in figure 6. There is slight increase in $h$ when $\varphi \leq$ 0.010 vol%. Then onwards till $\varphi$ =0.015, there is a steep rise in the curve and the maximum $h$ (902.68W/m²K) isobserved at $\varphi$ = 0.015. Subsequently h starts to drop till $\varphi$ = 0.02, but rises again with $\varphi$.

Nusselt number nearly follows the shape of h curve. Maximum Nu (4.67) was recorded when $\varphi$ reaches 0.015 vol%. Convective heat transfer enhancement was determined from these graphs and presented in Table 6.

The heat transfer improvement in the 0.015% Au nanofluid was around 7.00% to 8.64% of the base fluid[21].This justifies the 7.00% heat transfer enhancement in this simulation for the same volume fraction (Refer Table 6).

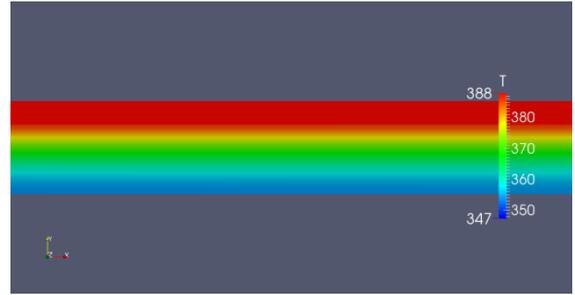

Figure 4: Temperature profile of 0.002% at steady state in fully developed laminar flow

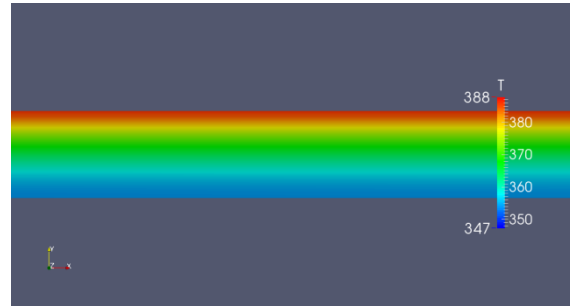

Figure 5: Temperature profile of 0.01 Vol% Au-water nanofluid at steady state in fully developed laminar flow

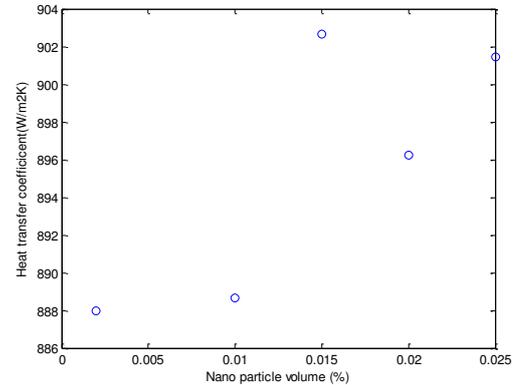

(a)

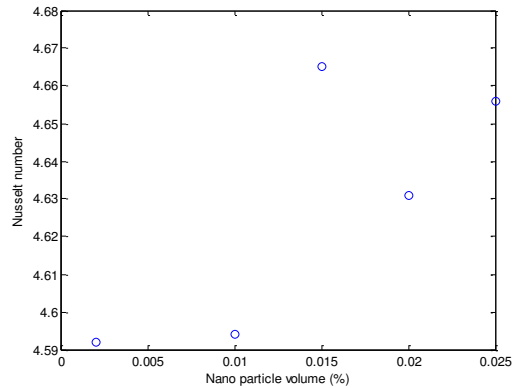

(b)

Figure 6: Relationship between Nanoparticle concentration, (a)Heat transfer coefficient, and (b) Nu number.

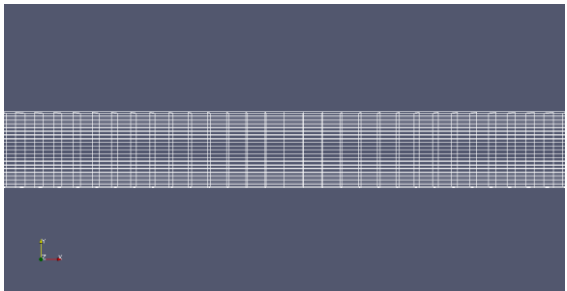

Figure 3: Computational mesh for the tube

**Table 6: Enhancement of Heat transfer coefficient with particle concentration**

| Au vol% | Heat transfer coefficient enhancement (%) |
|---|---|
| 0.002 | 5.34 |
| 0.01 | 5.37 |
| 0.015 | 7.00 |
| 0.020 | 6.21 |
| 0.025 | 6.80 |

**MDS Simulation of a nanoparticle**

Pair distribution function in figure 7 shows the variation of water density around the nanoparticle. This is a confirmation of liquid layering around the nanoparticle.

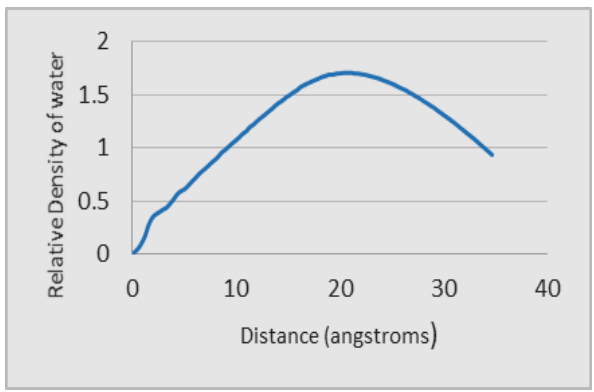

Figure 7: Distance vs. Relative Density of water around the nanoparticle

Furthermore, figures 8, 9 and 10 pictorially illustrate the distribution of water molecules in the liquid layer around the nanoparticle at various distances from the nanoparticle boundary.

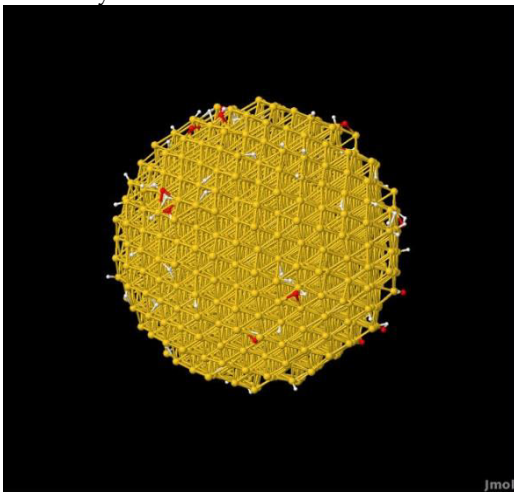

Figure 8: 1.5 Angstroms away from nanoparticle boundary

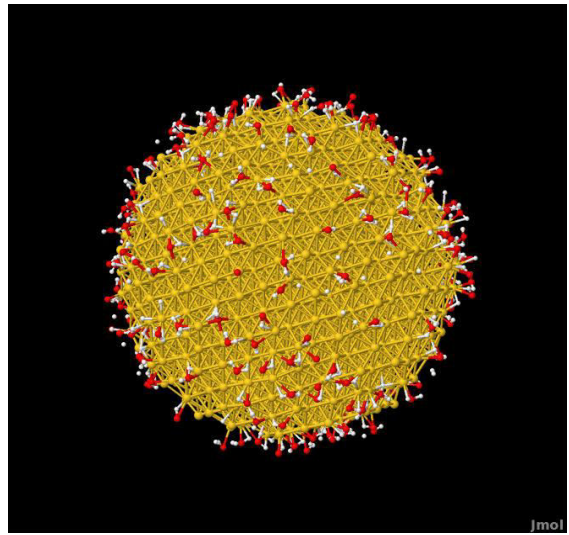

Figure 9: 2.5 Angstroms away from nanoparticle boundary

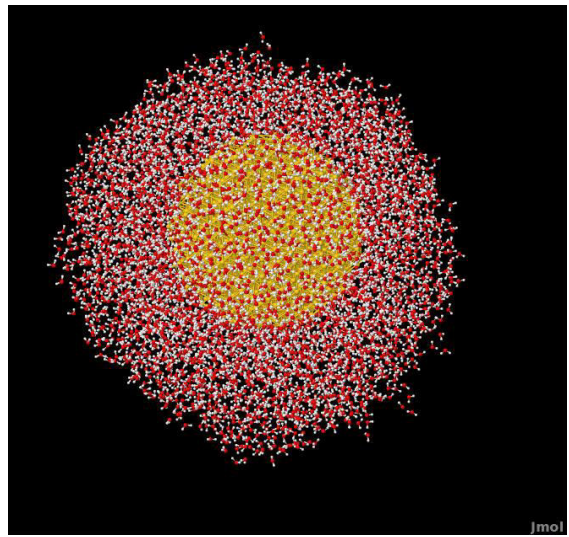

Figure 10: 34.5 Angstroms away from nanoparticle boundary

According to MDS results given above can be summarized as,
- There is no water layer close to the nanoparticle. This happen due to the repulsive power generated between water molecule and Au atoms.
- Beyond the 2.5 Angstroms from nanoparticle boundary it starts to form a surrounding water layer.
- Around 20 Angstroms (2nm) away from the nanoparticle boundary, water density reaches to a maximum value.

- Thereafter, 3.45nm away from the nanoparticle, relative density of water reaches 1.
- 

## 6. Conclusions

This paper examined the heat transfer in nanofluids. Gold nanoparticles were suspended in water and CFD simulations were conducted for fully developed laminar flow in a circular tube of diameter 6mm and length 100cm, under constant heat flux conditions. A clear increase of convective heat transfer was observed with increasing nanoparticle concentration. The maximum heat transfer coefficient of 902.68W/m²K occurred when the nanoparticle volume fraction was 0.015. This is equivalent to a 7.0% enhancement in comparison to base fluid water. One major cause for this increase could be the liquid layering around nanopartciles. To further verify this, OpenMD simulations were performed. Here 4nm Gold nanoparticles were submerged in SPCE water model and a 1709kg/m³ of maximum water density observed 2nm away from the nanoparticle boundary. This dense water layer may increase the thermal conductivity of the nanofluid and yields the observed enhancement.

Presented here are the preliminary results of a deeper investigation. Once completed, this work may lead to the manufacture of smart fluids for a wide range of engineering applications such as power plants, engines, HVAC systems and microelectronics.